\documentclass[letterpaper, 10 pt, conference]{ieeeconf}  

\IEEEoverridecommandlockouts                              

\usepackage{graphics} 
\usepackage{epsfig} 
\usepackage{mathptmx} 
\usepackage{times} 
\usepackage{amsmath} 
\usepackage{amssymb}  
\usepackage{amsfonts}
\usepackage{comment} 
\usepackage{tikz}
\usepackage{graphicx,xcolor}
\usepackage{subcaption}
\usepackage{bm}
\usepackage{algorithm}
\usepackage{algorithmic}
\usepackage{lipsum,adjustbox}
\usepackage{balance}

\newcommand{\startEW}{\color{black}}
\newcommand{\stopEW}{\color{black}}

\title{\LARGE \bf
Improving a Proportional Integral Controller with Reinforcement Learning on a Throttle Valve Benchmark
}

\author{Paul Daoudi$^{1,3}$, Bojan Mavkov$^{2}$, Bogdan Robu$^{3,*}$, Christophe Prieur$^{3}$, \\ Emmanuel Witrant$^{3}$, Merwan Barlier$^{1}$ and Ludovic Dos Santos$^{4}$
\thanks{$^{1}$ Paul Daoudi and Merwan Barlier are with the Noah's Ark Laboratory of Huawei Technologies, Paris, France. Email: {\tt\small paul.daoudi1@huawei.com} and {\tt\small merwan.barlier@huawei.com}}%
\thanks{$^{2}$ Bojan Mavkov is with Université Côte d'Azur, CNRS, I3S, Nice, France. Email:  {\tt\small bojan.mavkov@univ-cotedazur.fr}}%
\thanks{$^{3}$ Paul Daoudi, Bogdan Robu, Christophe Prieur and Emmanuel Witrant are with Univ. Grenoble Alpes, CNRS, Grenoble INP, GIPSA-lab, Grenoble, 38000, France Email: {\tt\small bogdan.robu@univ-grenoble-alpes.fr}, {\tt\small christophe.prieur@gipsa-lab.fr} and {\tt\small emmanuel.witrant@univ-grenoble-alpes.fr}}%
\thanks{$^{4}$ Ludovic Dos Santos is with Criteo AI Lab, Paris, France. Email: {\tt\small l.dossantos@criteo.com}}%
\thanks{$^{*}$ Corresponding author: Bogdan Robu.}
}

\begin{document}

\maketitle
\thispagestyle{empty}
\pagestyle{empty}

\begin{abstract}

This paper presents a learning-based control strategy for non-linear throttle valves with an asymmetric hysteresis, leading to a near-optimal controller without requiring any prior knowledge about the environment. We start with a carefully tuned Proportional Integrator (PI) controller and exploit the recent advances in \emph{Reinforcement Learning (RL) with Guides} to improve the closed-loop behavior by learning from the additional interactions with the valve. We test the proposed control method in various scenarios on three different valves, all highlighting the benefits of combining both PI and RL frameworks to improve control performance in non-linear stochastic systems. In all the experimental test cases, the resulting agent has a better sample efficiency than traditional RL agents and outperforms the PI controller.

\end{abstract}


\section{Introduction}

Throttle valves are essential components in various industrial processes such as chemical plants, oil refineries, and power generation. Accurate control of these valves is crucial for maintaining the optimal flow rate of fluids and thus ensuring the efficient functioning of the entire system. However, controlling throttle valves is complex due to their non-linear behavior, stochasticity, and asymmetric hysteresis. These factors make it challenging to design a controller that can handle the complexities of the valve system.

In this work, we focus on regulating a butterfly valve for car engines, which modulates the air supply to the combustion chamber by adjusting the disc's rotation angle. Given the complexities \startEW induced \stopEW by static friction and the \startEW non-linearities of the \stopEW dynamics, the control community has explored various advanced strategies. \startEW For example, non-linear control approaches include \stopEW an adaptive pulse control method leveraging a non-linear dynamic model accounting for friction and aerodynamic torque \cite{de2001adaptive}, a discrete-time sliding mode control for robust tracking \cite{ozguner2001discrete}, and a non-linear approach combining a Proportional-Integral-Derivative (PID) controller with a feedback compensator \cite{deur2004electronic}. Additionally, adaptive control techniques \cite{pavkovic2006adaptive,jiao2014adaptive,witrant2023teaching}, Linear Parameter-Varying (LPV) modeling, and mixed constrained $H_2/H_{\infty}$ control strategies \cite{zhang2014lpv} have been proposed. Several works also exploit deep learning techniques. For instance, \cite{baric2005neural} presents a neural network-based sliding mode controller for electronic throttle valves. A control strategy using \startEW RL \stopEW algorithms is advocated in \cite{siraskar2021reinforcement} where the valve is controlled using the Deep Deterministic Policy Gradient (DDPG) \cite{LillicrapHPHETS15} algorithm. However, the application of the algorithm is limited to simulations.

Our contribution extends the existing body of work by examining the combination of \startEW a classical PI \stopEW  control strategy with \startEW RL \stopEW for throttle valve regulation. The RL framework is popular due to its ability to solve complex non-linear control problems without requiring an explicit model of the system \cite{sutton2018reinforcement,arulkumaran2017deep}. However, despite some exceptions such as balloon navigation \cite{bellemare2020autonomous} and plasma control in Tokamaks \cite{degrave2022magnetic}, RL has only been applied in simulated systems \cite{koch2019reinforcement, tunyasuvunakool2020dm_control}, including the aforementioned work \cite{siraskar2021reinforcement}. One major obstacle in directly applying RL to real-world systems is the need for a large amount of data and repetitive experiments to learn a good policy, as RL agents start in an unknown environment with no prior information available \cite{Dulac_ArnoldLML21}.

\begin{figure*}[t!]
    \centering
    \includegraphics[width=0.99\linewidth]{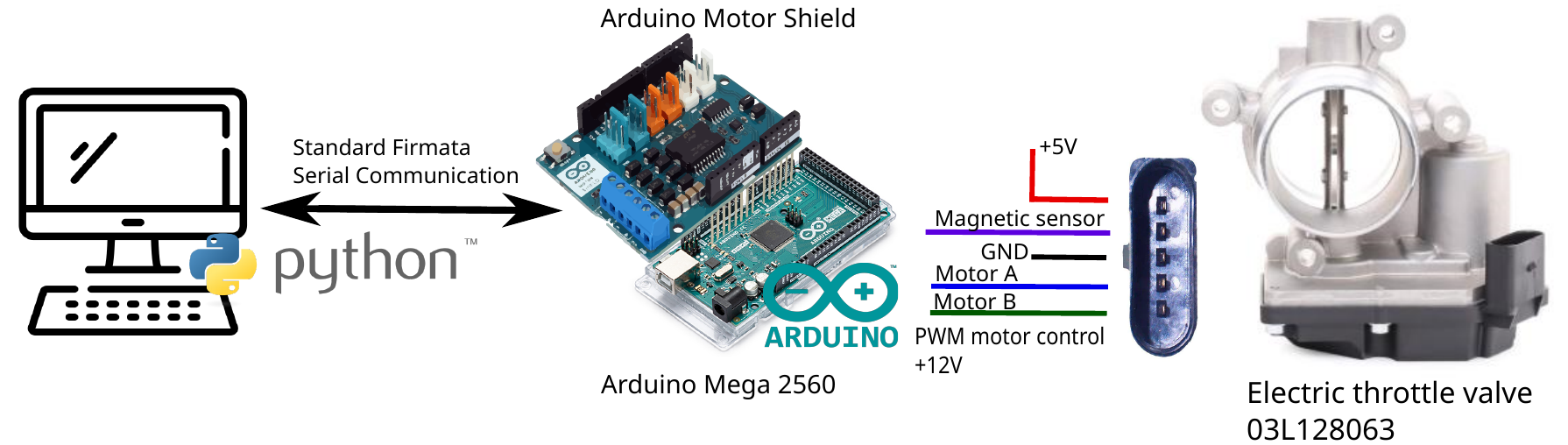}
    \caption{Experimental test bench: hardware configuration and
wiring of the electric throttle valve.}
    \label{fig:arduino_board}
\end{figure*}

Only a few recent works combine Control Theory (CT) and RL to reduce the number of data required by RL agents to learn a good policy \cite{recht2019tour}.

The approaches of \cite{gillen2020combining} and \cite{zoboli2021reinforcement} propose a switching mechanism between an LQR controller and an RL agent, testing these approaches in simulated environments. \startEW Some approaches \stopEW embed a feedback controller within an RL policy to enhance sample efficiency \cite{johannink2019residual,zhang2023aiding}. However, these methods do so without restraining the search space for the control input, nor without modifying the learning of the $Q$-function, which may be essential to avoid extrapolation errors \cite{fujimoto2019off}. Similar to these previous methods, we aim to combine the strengths of CT and RL to build a controller for the throttle valve. For this, we adapt the recent area of Reinforcement Learning with Guides \cite{zimmer2014teacher, agarwal2022reincarnating, daoudi2023enhancing} to our setting, resulting in an algorithm that \startEW leverages \stopEW a Proportional Integral (PI) controller to guide the RL agent. By reducing the search space for learning, the PI guide minimizes the data requirements and accelerates the learning process. We stress that our method is fundamentally different than \startEW those proposed by \stopEW \cite{sedighizadeh2008adaptive,carlucho2020adaptive, lawrence2020optimal}\startEW , which \stopEW use RL to tune the gains of a feedback PI controller. Here, the PI controller is fixed and is used to guide the RL agent to improve its sample efficiency.

We conduct empirical validation of our approach on the throttle valve by testing the PI controller, a state-of-the-art RL agent, and their combination under a variety of industry-relevant use cases. This empirical validation is repeated on \startEW three \stopEW valves having the same commercial reference but possessing slightly different physical properties. Our results demonstrate that the combination of PI and RL produces an effective agent in all settings, achieving near-optimal control with fewer data samples compared to using a traditional RL agent alone.

The paper is structured as follows. Section~\ref{sec:throttlesystem} \startEW describes \stopEW the experimental setup and outlines the physical properties of the considered throttle valves. Section~\ref{sec:traditional_techniques} formulates our objectives and introduces two control methods that are commonly used for similar systems. Section~\ref{sec:combination} proposes our hybrid control algorithm, blending the strengths of classical and RL methodologies. Finally, Section~\ref{sec:experiments} presents our experimental findings, \startEW evaluating the \stopEW strengths of the proposed controllers \startEW according to different criteria\stopEW.

\section{The throttle valve system}\label{sec:throttlesystem}

This section presents a detailed overview of our experimental setup for controlling throttle valves, explores their physical properties, and formalizes the objective.

\subsection{Experimental setup}

The throttle valves investigated have the commercial reference 03L128063 \startEW and included in an experimental test bench for control as detailled in \stopEW \cite{witrant2023teaching}. They feature a rotational spring on the shaft of the valve plate, exerting a counteractive torque against the motor's torque to control the plate's angular position. In this study, we consider $3$ valves coming from the above reference.

To regulate the opening angle, a Pulse Width Modulation (PWM) ranging from $0$ to $100\%$ is generated and modulates the input voltage from $0$ to \startEW 12 \stopEW Volts. The valves are all equipped with a magnetic angle sensor. We connect each valve to an Arduino Mega 2560® that controls the input and allows monitoring and \startEW analyzing \stopEW the system performance. The motor control is enabled by the dual full-bridge driver \emph{Arduino Motor Shield} attached to the Arduino board. In addition, each Arduino is connected to a Python interface that is necessary to build and train the RL agent. For all \startEW the \stopEW valves, \startEW the sampling time for data acquisition and feedback control is \stopEW $50$ milli-seconds. The scheme of the experimental test bench is \startEW presented \stopEW in Figure~\ref{fig:arduino_board}.

\subsection{A non-linear stochastic system}

\begin{figure*}[t!] 
\begin{subfigure}{.33\textwidth}
  \centering
  \includegraphics[width=.99\linewidth]{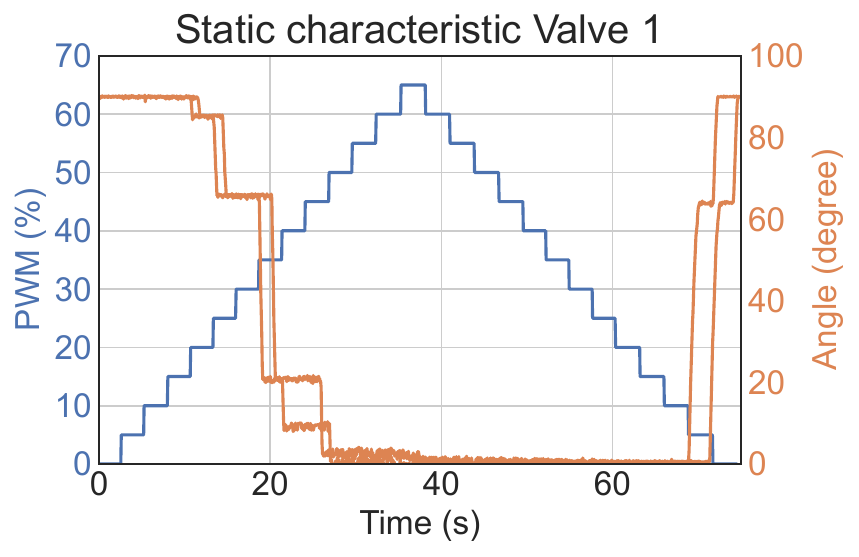}
  \label{fig:sophie_time_responses}
\end{subfigure}%
\begin{subfigure}{.33\textwidth}
  \centering
  \includegraphics[width=.99\linewidth]{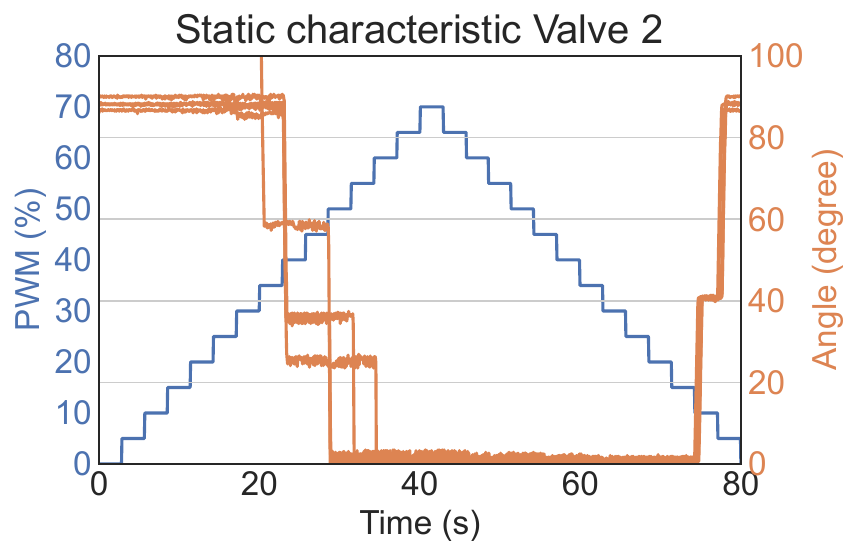}
  \label{fig:adama_time_responses}
\end{subfigure}%
\begin{subfigure}{.33\textwidth}
  \centering
  \includegraphics[width=.99\linewidth]{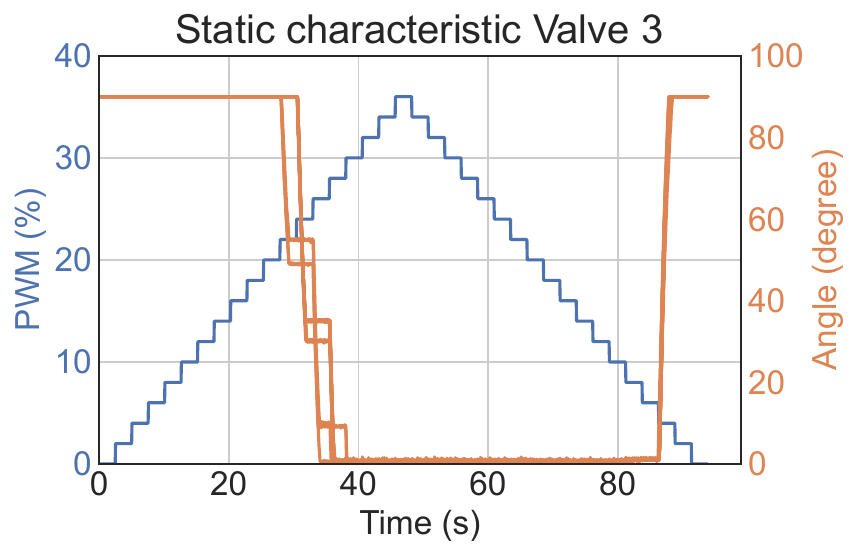}
  \label{fig:paul_time_responses}
\end{subfigure}
\begin{subfigure}{.33\textwidth}
  \centering
  \includegraphics[width=.99\linewidth]{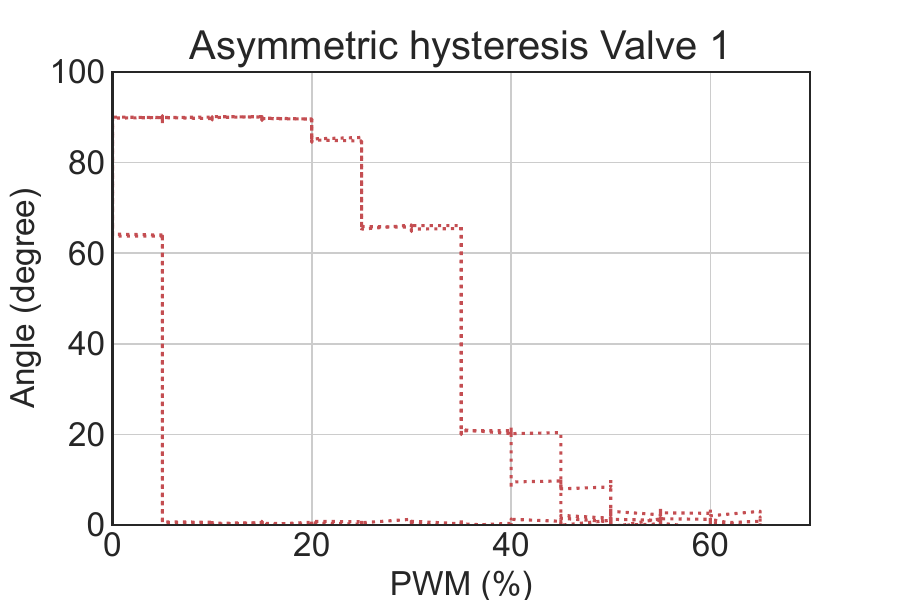}
  \label{fig:sophie_hysteresis}
\end{subfigure}
\begin{subfigure}{.33\textwidth}
  \centering
  \includegraphics[width=.99\linewidth]{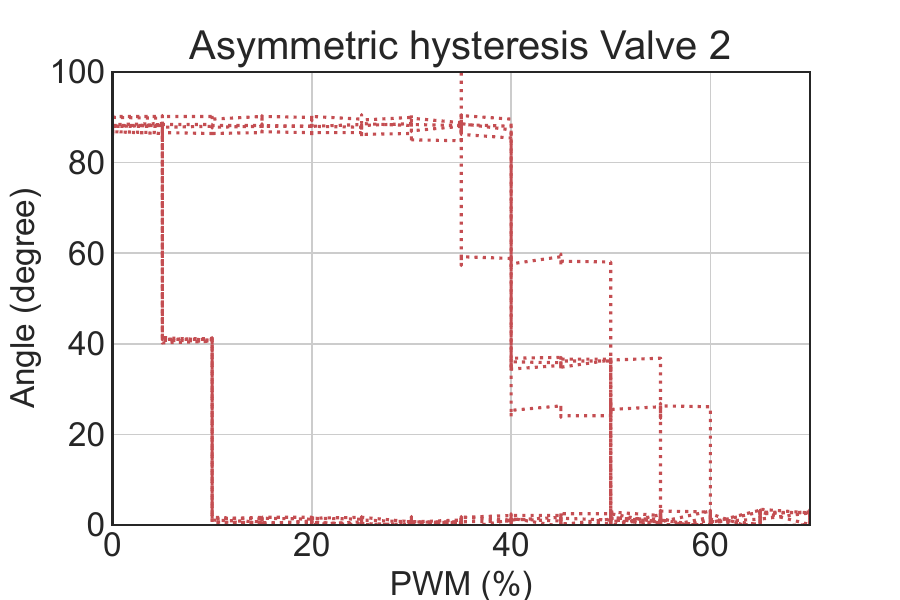}
  \label{fig:adama_hysteresis}
\end{subfigure}
\begin{subfigure}{.33\textwidth}
  \centering
  \includegraphics[width=.99\linewidth]{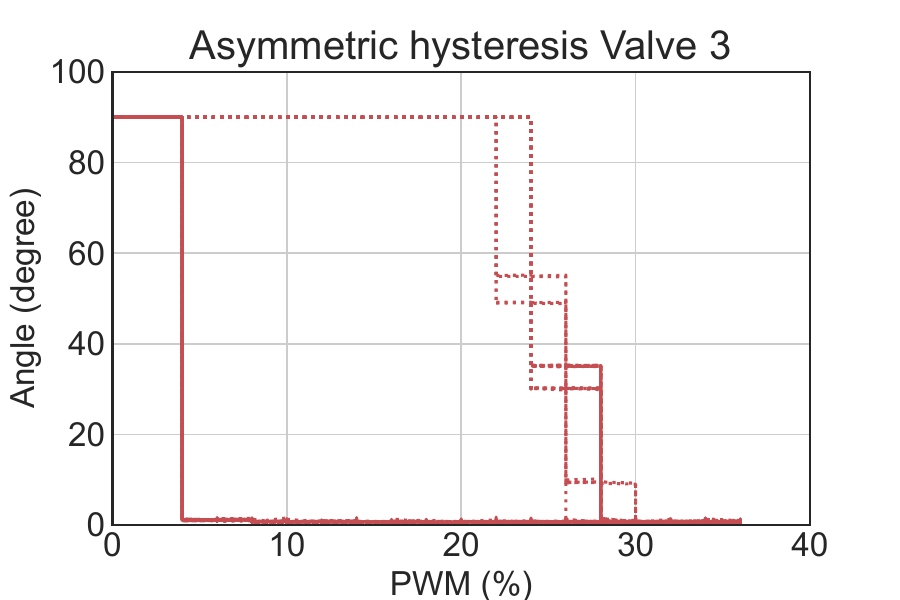}
  \label{fig:paul_hysteresis}
\end{subfigure}
\caption{Steady state analysis for the $3$ valves.} 
\label{fig:time_responses}
\end{figure*}

This type of valve is particularly known for its non-linear stochastic dynamics with asymmetric hysteresis, which we confirm in the following experiment. We select \startEW an input range \stopEW  from $0\%$ to $\text{PWM}_{\text{MAX}}\%$ and \startEW generate an increasing then decreasing sequence of steps covering this range, by increments of $5\%$\stopEW, and apply it to the valve. Despite the valves coming from the same commercial reference, we notice \startEW that \stopEW the required value $\text{PWM}_{\text{MAX}}$ to turn the valve's angular plate to $0$ degrees is different \startEW from one valve to another \stopEW  \cite{witrant2023teaching}. It is $\text{PWM}_{\text{MAX}} = 0.65\%$ for Valve $1$, $\text{PWM}_{\text{MAX}} = 0.70\%$ for Valve $2$ and $\text{PWM}_{\text{MAX}} = 0.45\%$ for Valve $3$. The experiment is repeated $5$ times, and we present the resulting data in Figure \ref{fig:time_responses}. An asymmetric hysteresis is observed for all valves, along with the stochastic nature of their dynamics. \startEW For a given PWM input\stopEW, the resulting angles vary significantly between each trial.

These results \startEW underlines \stopEW the necessity of utilizing advanced control strategies to effectively manage the intricacies of the dynamics of the valve. In addition, it highlights that each valve requires a slightly different controller to work.

\subsection{\startEW Problem statement and notations \stopEW}

Considering the stochastic nature of the valves' dynamics, the problem is modeled as a discounted Markov Decision Process (MDP) $\left( \mathcal{X}, \mathcal{U}, c, P, \gamma \right)$. Let $\Delta(\cdot)$ be the simplex on any given space ($\cdot$), and define $c : \mathcal{X} \times \mathcal{U} \to \mathbb{R}$ the cost function, $ P : \mathcal{X} \times \mathcal{U} \to \Delta(\mathcal{X}) $ the transition probabilities, and $\gamma\in \left[0,1\right)$ the discount factor. At each time step $t$, the agent receives an observation $x_t\in\mathcal{X}$ and selects a control $u_t\in\mathcal{U}$. The system is then updated with $x_{t+1}\sim P \left( \cdot | x_t, u_t \right)$.

In this formalism, the objective is to find the optimal controller $\pi^* : \mathcal{X} \to \Delta(\mathcal{U})$ that minimizes the expected discounted cumulative cost $V^\pi(x) = \mathbb{E}_{P} \left[ \sum_{t=0}^\infty \gamma^t c(x_t, u_t) | x_0=x, u_{t} \sim\pi(\cdot|x_t) \right]$ for each observation $x\in\mathcal{X}$ with as few data samples as possible. The function $V^\pi$ is known as the Value function of policy $\pi$, and we define the associated $Q$-value function as $Q^\pi(x, u) = \mathbb{E}_{P} \left[ \sum_{t=0}^\infty \gamma^t c(x_t, u_t) | x_0=x, u_0=u, u_{t} \sim \pi(\cdot|x_t) \right] $.

The goal of this work is to develop a throttle valve controller $\pi: \mathcal{X} \to \Delta(\mathcal{U})$ that sets the angle of the valve's plate to a chosen angle from any initial position. The controller must not only perform optimally in nominal operating conditions by minimizing cumulative costs but also demonstrate robustness under various real-world scenarios. Specifically, the controller must be able to reject external disturbances that may affect both the input and output signals, while exhibiting minimal overshoot. Bearing this information in mind, we associate the different components of the MDP with the valve. \startEW At time $t$, \stopEW let $\alpha_{t}$ be the angle of the valve's plate and $\alpha_{\text{ref}_t}$ be the \startEW (desired) reference \stopEW angle, and:
\begin{itemize}
    \item the observation $x_t$ at time $t$ is composed of the reference angle $\alpha_{\text{ref}_t}$, the previous and current angles $\alpha_{t-1}$ and $\alpha_t$, and the previous control input $u_{t-1}$;
    \item the control $u_t$ at time $t$ is the PWM ($\%$) input. To avoid saturation effects, we set it to $\mathcal{U} = [0, 0.8]$ for Valve $1$ and $2$, and to $\mathcal{U} = [0, 0.6]$ for Valve $3$;
    \item the objective is to find a controller $\pi: \mathcal{X} \to \Delta(\mathcal{U})$ that sets the angle of the valve's plate to a chosen angle from any initial position. Hence, the cost function $c$ is set to $c(x_t, u_t) = \left\| \alpha_t - \alpha_{\text{ref}_t} \right\|_2$;
    \item the controller must be optimal in all states, so the discount factor $\gamma$ is set to the high value of $0.99$;
    \item the transition probabilities $P$ are unknown;
    \item to enhance the convergence of RL algorithms, the system is made episodic, with each episode lasting for 100 time-steps, which represents $5$ seconds. \startEW During \stopEW each episode, a single random reference angle $\alpha_{\text{ref}_t}$ is generated, and the primary objective is to set the throttle valve's position to this specific angle.
\end{itemize}
\section{Traditional techniques}\label{sec:traditional_techniques}

In this section, we detail two traditional techniques that will be used to achieve our objective: regulate the opening of the throttle valve while being robust to the inherent noise and the stochasticity of the dynamics.  

\subsection{\startEW Discrete time \stopEW PI Controller}

The PI controller is commonly used in control systems design due to its simplicity and effectiveness in a wide range of systems. It can be expressed \startEW in discrete time, using the notations introduced in the previous section, \stopEW as:
\begin{equation}
    \pi_{\text{PI}}(x_t) = u_{t-1} - r_0 \alpha_t - r_1 \alpha_{t-1} + (r_0 + r_1) \alpha_{\text{ref}_t}.
\end{equation}

In order to properly tune the gains for this system, we followed the steps proposed by the previous work \cite{witrant2023teaching}, detailed below. This work found that the following auto-regressive model
\begin{equation} \label{eq:auto_model}
    \alpha_{t} = a \alpha_{t-1} + b_1 u_{t-1} + b_2 u_{t-2}
\end{equation}
represents the dynamics of any valve from the considered commercial reference \startEW sufficiently well for control purposes\stopEW. The model parameters $a$ and $b$ are tuned to fit the first order model \startEW using \stopEW data describing the response of the valve from a \startEW sufficiently rich input signal (from the persistency of excitation point of view)\stopEW, generated by a $1022$-\startEW long \stopEW Pseudo Random Binary Sequence (PRBS) centered at $16\%$. Then, $r_0$ and $r_1$ are chosen such that the damping ratio $\zeta$ is set to $1$ and the rising time $t_R$ \startEW is \stopEW $0.8$ seconds. 

\startEW For the valves considered in this paper\stopEW, this experiment leads to the values of ($a=0.78, \ b_1=-0.18, \ b_2=-0.23, \ r_0 = -2.28, \ r_1 = 1.83$) for Valve $1$, ($a=0.74, \ b_1=-0.25, \ b_2=-0.41, \ r_0 = -1.31, \ r_1 = 1.01$) for Valve $2$, and ($a=0.83, \ b_1=-0.11, \ b_2=-0.23, \ r_0 = -2.33, \ r_1 = 1.96$) for Valve $3$.

Note that the purpose here is not to design the best possible PI controller but to show the advantages of combining both control and machine learning techniques, \startEW in comparison with \stopEW using control or reinforcement learning techniques separately.

\subsection{Approximate Policy Iteration}

To have a reliable effective controller for the valve, we propose to use a traditional \startEW RL \stopEW agent that uses the \emph{Approximate Policy Iteration} scheme \cite{bertsekas2011approximate}.

Most RL algorithms rely on the Bellman operator, defined as $\mathcal{B}^\pi\left[Q\right](x,u) = c(x,u) + \gamma \mathbb{E}_{x'\sim P(\cdot|x,u), u'\sim \pi(\cdot|x')} \left[ Q(x',u') \right]$. This operator is a $\gamma$-contraction, so iteratively applying it to any function $Q$ converges to its unique fixed point $Q^\pi$. However, since the transition probabilities $P$ are unknown, the empirical Bellman operator $\hat{\mathcal{B}}$ that uses interactions with the environment to estimate this expectation is used instead.

In the Approximate Policy Iteration framework, both the policy $\pi_{\text{RL}}$ and the $ Q$ functions are parametrized with the respective weights $\theta\in\Theta$ and $\omega\in\Omega$, \startEW and are thus \stopEW denoted as $\pi_{\text{RL}}^\theta$ and $Q^\omega$. Both estimates will work together until the policy converges to a near-optimal one. More \startEW precisely\stopEW, at each epoch $k$, the agent collects data with the current policy $\pi_{\text{RL}}^{\theta_k}$, evaluates its associated $Q$-function via Approximate Policy Evaluation ~(\ref{eq:policy_eval}) and improves its policy through Approximate Policy Improvement~(\ref{eq:policy_improvement}). Given a dataset $\mathcal{D} = \{ (x_i, u_i, c_i, x_{i+1})_{i=1}^N \}$, $\hat{\mathbb{E}}$ the empirical expectation of the observation-control pair $(x, u)$ induced by the data set $\mathcal{D}$, the scheme is formalized as: 
\begin{align}
    & Q^{\omega_{{k+1}}} \leftarrow \arg\min_{\omega\in\Omega} \,\hat{\mathbb{E}} \left[ \left( Q^\omega - \hat{\mathcal{B}}^{\pi^{\theta_k}_{\text{RL}}}\left[ Q^{{\omega}_k} \right] \right)^2 \right] \tag{APE},\label{eq:policy_eval} \\
    & \pi^{\theta_{k+1}}_{\text{RL}} \leftarrow \arg\min_{\theta\in\Theta} \, \hat{\mathbb{E}}_{x\sim\mathcal{D}, u\sim\pi^\theta_{RL}} \left[ 
    Q^{\omega_{k+1}}(x,u)
    \right]\tag{API}\label{eq:policy_improvement}.
\end{align}

This cycle of evaluation and improvement is commonly solved approximately with gradient descent. This framework, not requiring prior knowledge from the environment beyond interaction data, can be applied as such to any system as long as it possesses the Markov property. Nonetheless, the absence of initial information makes it difficult to collect meaningful data at the early stages of learning. This leads to poor estimates of the $Q$ function and the policy. As a result, agents necessitate extensive interactions to build accurate estimates and derive a near-optimal policy.

TD3 or Twin Delayed Deep Deterministic Policy Gradient \cite{fujimoto2018addressing}, is a leading algorithm in this framework. Building upon the foundation laid by DDPG \cite{LillicrapHPHETS15}, it estimates the $Q$-function using two neural networks and introduces delayed policy updates to enhance the stability of learning. 

\section{Combining both methods} \label{sec:combination}

\begin{figure*}[t!]
    \centering
    \includegraphics[width=\linewidth]{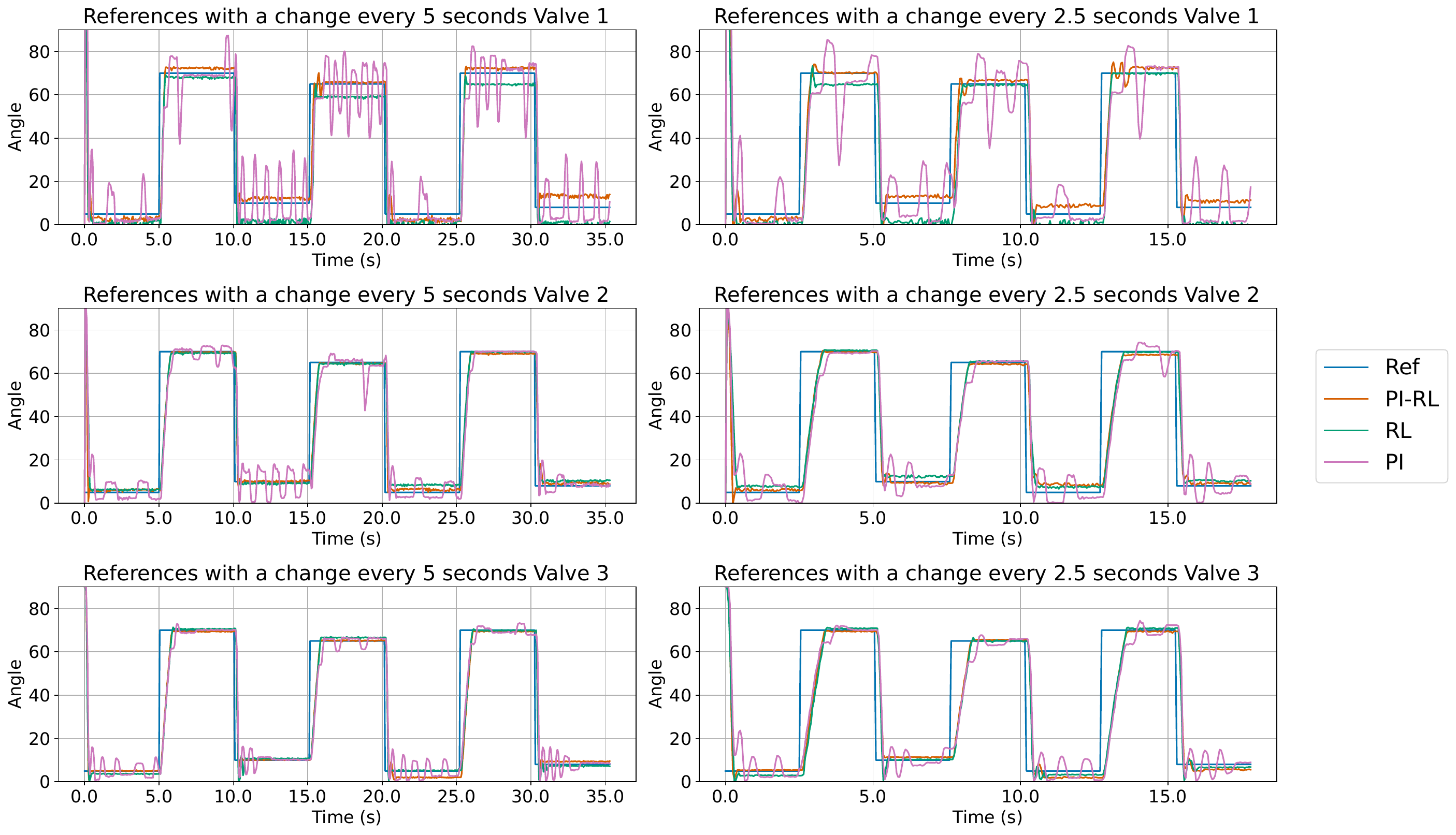}
    \caption{Comparison of agents under different scenarios with access to real outputs. Both figures show the evolution of angles over time with a reference change every $X$ seconds, set to 5 seconds for the left figures and 2.5 seconds for the right figures.
    } 
    \label{fig:speed_comparison}
\end{figure*}

On the one hand, the PI controller is easy to build and requires a minimal amount of data, but suffers from sub-optimality on non-linear systems due to its linear dependencies. On the other hand, RL agents often achieve near-optimality on complex systems but require a significant amount of data. Two primary reasons for this requirement are the poor initialization of the policy and the exhaustive search for the control $u$ across the whole control space $\mathcal{U}$ at each time step.

To combine the advantages of both methods, we draw inspiration from the recent framework \startEW of \stopEW \emph{Reinforcement Learning with Guides} \cite{zimmer2014teacher} and adapt one of the most recent and efficient techniques to our context: Perturbation Action Guided (PAG) \cite{daoudi2023enhancing}. For clarity purposes, we refer to the adaptation of PAG to our setting by PI-RL. 

In this approach, to circumvent searching the entire control space $\mathcal{U}$ for the optimal control, the PI controller is incorporated to guide the training process. This not only reduces the search space but also steers the RL agent towards favorable parts of the environment, directly improving exploration and the quality of the gathered data. To do so, the PI-RL policy is centered around the PI controller and learns a perturbation $\xi^\phi_{\text{RL}}$, with $\phi\in\Xi$ to guide it towards better actions. Formally, the new policy is written:
\begin{equation} \label{eq:policy_pag}
    \pi_{\text{PI-RL}}^{\phi}(\cdot\vert x) = \pi_{\text{PI}}(x) + \xi^{\phi}_{\text{RL}}(\cdot \vert x).
\end{equation}

Unlike the traditional RL policy $\pi_{\text{RL}}^\theta$ that provides a control input $u$, the perturbation $\xi^\phi_{\text{RL}}$ aims solely to enhance the PI controller's performance. Thus, rather than defining the perturbation $\xi^{\phi}_{\text{RL}}$ over the entire control space $\mathcal{U}$, it is constrained to a smaller subset $S(\mathcal{U})$. This confines the search for control to a vicinity around the PI controller, where the optimal control is more likely to be found. In practice, to reduce the search space (originally $[0, 0.6]$ or $[0, 0.8]$) while keeping the policy differentiable, a multiplicative factor $\eta\in [0, 1]$ is introduced that scales the control input by $\eta$.

The perturbation is learned similarly to \startEW a control input \stopEW in traditional RL, that is by minimizing its associated $Q^\omega$ function. \startEW Note that \stopEW the expectation in the Bellman operator is with respect to $\pi^{\phi_k}_{\text{PI-RL}}$, not $\pi_{\theta_k}$. \startEW This \stopEW avoids the overestimating problem that may be encountered when the $Q$-function estimates observation-control pairs that are not described by the data set $\mathcal{D}$ \cite{fujimoto2018addressing,daoudi2023enhancing}. The process becomes the PI-Approximate Policy Iteration, where the APE-API steps are modified to be PI-APE and PI-API, defined as:
\begin{align}
    & Q^{\omega_{{k+1}}} \leftarrow \arg\min_{\omega\in\Omega} \,\hat{\mathbb{E}} \left[ \left( Q^\omega - \hat{\mathcal{B}}^{\pi^{\phi_k}_{\text{PI-RL}}}\left[ Q^{{\omega}_k} \right] \right)^2 \right] \tag{PI-APE},\label{eq:pag_policy_eval} \\
    & \xi^{\phi_{k+1}}_{\text{RL}} \leftarrow \arg\min_{\phi\in\Xi} \, \hat{\mathbb{E}}_{x\sim\mathcal{D}, u\sim\pi^\phi_{\text{PI-RL}}} \left[ 
    Q^{\omega_{k+1}}(x,u)
    \right]\tag{PI-API}\label{eq:pag_policy_improvement}.
\end{align}

Similarly to the traditional RL agent, both the $Q$-function and the perturbation $\xi^\phi_{\text{RL}}$ are parametrized with neural networks that are learned with gradient descent. The scaling term $\eta$ is set to $0.5$ for both valves reducing by two the search space. The subspace of $\mathcal{U}$ is hence set to $S(\mathcal{U}) = [0, 0.4]$ for the first two valves and to $S(\mathcal{U}) = [0, 0.3]$ for the last valve.

The whole algorithm can be found in Pseudo-Code \ref{alg:pag_alg}.

\begin{algorithm}[H]
\begin{algorithmic}
\STATE Select damping ratio $\zeta$, rising time $t_R$ and $\Phi$
\STATE Create PI controller 
\STATE Initialize $Q^{\omega_0}$ and $\xi^{\phi_0}_{\text{RL}}$
\FOR {$k \in (0, \dots, K)$} 
    \STATE Gather data with $\pi^{\phi_k}_{\text{PI-RL}}$ from Eq.~(\ref{eq:policy_pag})
    \STATE Add data to the replay buffer $\mathcal{D}$
    \STATE Sample a batch from $\mathcal{D}$
    \STATE Update $Q^{\omega_{k+1}}$ with gradient descent on Eq.~(\ref{eq:pag_policy_eval}) 
    \STATE Update $\xi^{k+1}_\phi$ with gradient descent on Eq.~(\ref{eq:pag_policy_improvement})
\ENDFOR
\end{algorithmic}
\caption{PI-RL}
\label{alg:pag_alg}
\end{algorithm}

\section{Comparison of the different controllers \startEW on the experimental test benches \stopEW} \label{sec:experiments}

\begin{figure*}[ht!]
    \centering
    \includegraphics[width=0.99\linewidth]{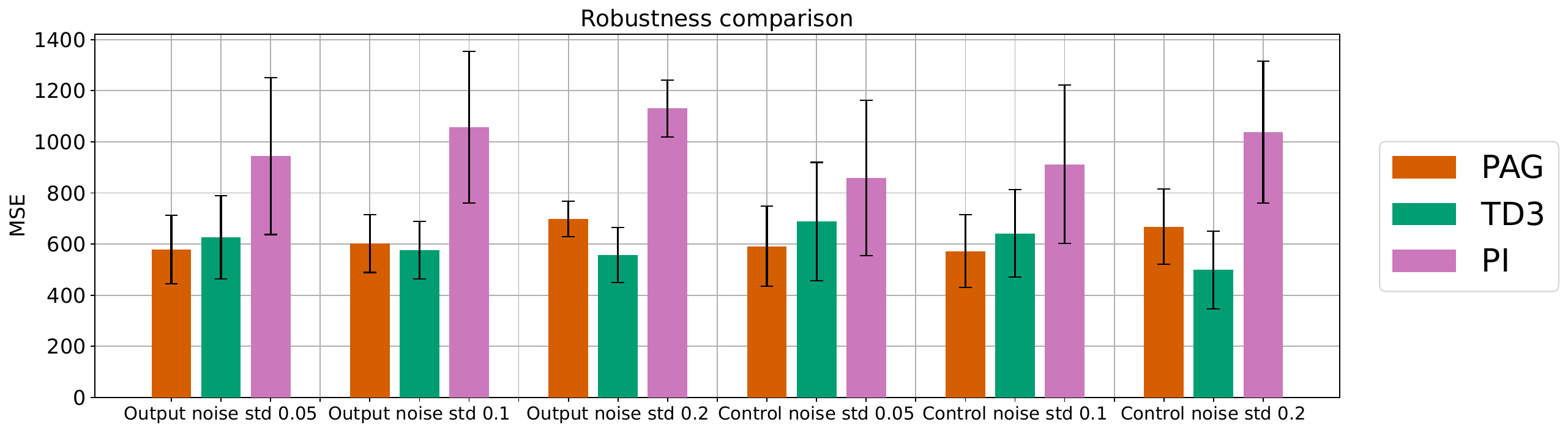}
    \caption{Comparison of agents under noisy outputs or controls with varying standard deviations. Each bar represents the mean and standard deviation of the Mean-Squared-Error (MSE) between the reference and the angle across the $3$ valves.}
    \label{fig:robustness_comparison}
\end{figure*}

We conduct a series of experiments on various tasks, including tracking a reference signal with sufficient time to adjust, adapting to quick reference changes, and demonstrating robustness with respect to noise in the control input or in the output observation.

To assess the effectiveness of the different controllers, we use the Mean Square Error (MSE) between the current angle and the objective. It is formally defined with sequences $\alpha = (\alpha_0, \alpha_1, \dots)$ and $\alpha_{\text{ref}} = (\alpha_{\text{ref}_0}, \alpha_{\text{ref}_1}, \dots)$ as:
\begin{equation}
    MSE(\alpha, \alpha_{\text{ref}}) = \left\| \alpha - \alpha_{\text{ref}} \right\|_2.
\end{equation}
This metric seems appropriate as it encompasses all of the costs during one scenario.


In all our experiments, both RL-based algorithms use TD3. The $Q$-functions and the policies $\pi^\theta_{\text{RL}}$ and $\xi^\theta_{\text{RL}}$ are parametrized with neural networks with $2$ hidden layers of size $64$ and use the ReLU activation function.

\subsection{Nominal comparison and reaction to quick reference changes}

We compare the performance of all agents for tracking a reference signal through a series of experiments. First, we perform a comparison when the angles change every $5$ seconds, and one with quicker reference changes every $2.5$ seconds. All results are summarized in Figure \ref{fig:speed_comparison}. 


\paragraph{PI controller} In all scenarios, we observe that the PI controller, while efficient, encounters difficulties in accurately tracking the reference signal. The controller requires time to accumulate errors before adjusting to the reference. Due to the system's complexity \startEW and the dry friction effect\stopEW, it may cause the valve's plate to overshoot the target. The PI controller would then need additional time to adjust, and potentially select overly aggressive inputs once more. This results in oscillations around the reference angle, particularly evident in the first valve. The other valves also present such oscillations, especially for low-angle references. This reveals the need for more sophisticated controllers and not solely relying on a linear policy.


\paragraph{RL-based agents} On the other hand, both RL-based agents demonstrate remarkable success in precisely tracking the reference signal in all scenarios, including the most challenging ones. For example, they can track all reference values with a change every 2.5 seconds, which is twice as fast as the rate they were trained for. This level of performance is also observed for low-angle references that proved to be challenging for the PI controller. The controls provided by the RL-based agents are stable in all settings, with no instances of unstable behavior observed. In some cases, a slight tracking bias is observed where the agents reach a point close to but not precisely at the reference. This can be attributed to the RL agents' goal of minimizing the $Q$ function, which is an expectation over the system's stochastic dynamics. To address this, a more appropriate RL formulation may be needed, which considers higher-order moments of the cumulative future cost distribution or steady-state errors in a refined way. We leave this for future work. Despite these minor biases, both RL-based agents demonstrate a strong performance in all tasks for all valves.

\subsection{Robustness to external noise}

We also conduct experiments to assess the robustness of the controllers in the presence of noise, both in the output and in the control. To this end, we test the controllers under varying noise levels and analyze their performance.

To better visualize the effects of noise on the controllers' performance, we plot in Figure \ref{fig:robustness_comparison} the Mean Squared Error (MSE) rather than the signals and the reference as in the previous section.

In comparison to the PI controller, both RL-based agents demonstrate superior performance over the PI controller in all scenarios. These RL-based agents have approximately the same performance in this setting, which is expected given both of them have been trained until convergence. It is also essential to note that, under specific conditions of noise, e.g., with an output or control noise of std $0.2$, the PI-RL agent exhibits a higher loss than the traditional RL agent. This result may be attributed to the fact that the PI-RL agent is centered around the PI controller which is inadequate in these scenarios: the reduced search space may not be sufficient to yield an optimal controller. Nevertheless, the perturbation $\xi_{RL}^\phi$ can improve the PI controller up to a certain point.

In summary, our robustness analysis reveals that both RL-based agents perform well in noisy environments while the PI controller's performance declines more rapidly. This declination may affect the PI-RL agent as the search space is too narrow to recover a near-optimal policy. These results suggest that the PI-RL agent offers promising results in low-to-moderate noise environments, where the PI controller can still provide a relevant control signal. However, in more challenging noise conditions where the PI controller's performance is limited, traditional RL may be better suited to provide a more robust and effective controller.

\subsection{Sample efficiency}

\begin{figure}[t!]
\centering
\includegraphics[width=.75\linewidth]{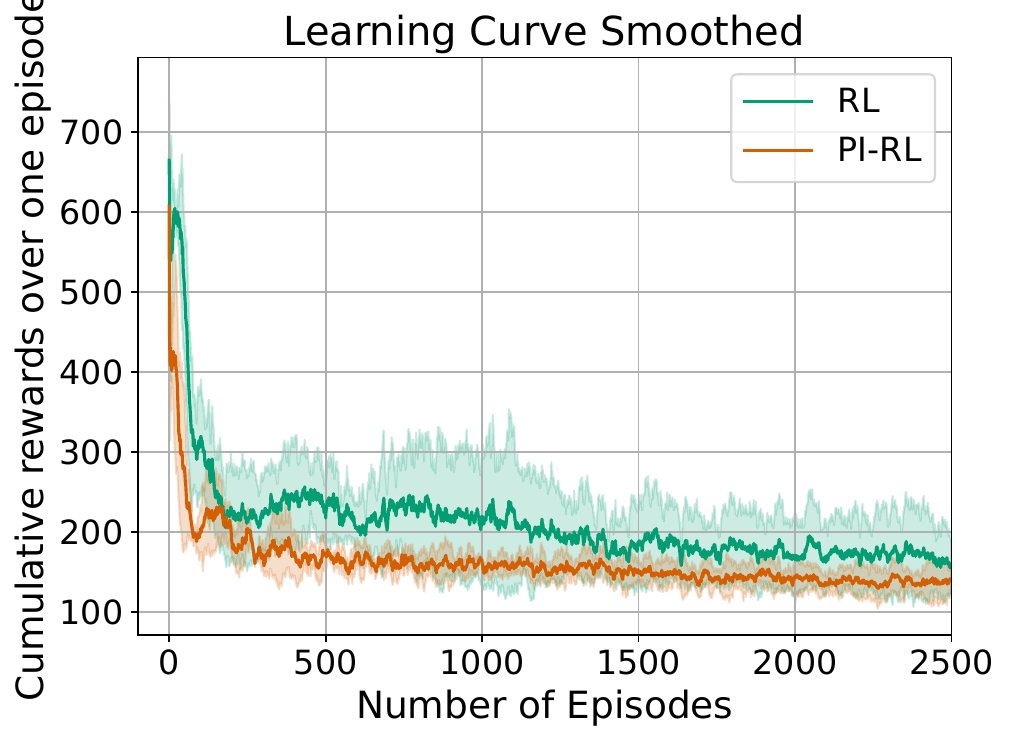}
    \caption{Learning curves of the different RL-based agents averaged on the $3$ valves. Agents are both trained with $4$ CPUs during $2500$ episodes, representing $3$ hours. }
    \label{fig:learning_curves}
\end{figure}

As previously shown, both the PI-RL and classical RL agents achieve similar performances in controlling the throttle valve. One important advantage of the combined PI and RL approach over the classical RL approach is that it requires fewer data to learn even though the PI controller may not be the optimal one. This can be seen in Figure~\ref{fig:learning_curves}, where we plot the cumulative costs over the training process for both controllers averaged over the $3$ different valves.

We observe that the PI-RL approach achieves a low cumulative cost faster than the RL approach, indicating that it was able to learn the control task more efficiently. This is because PI-RL leverages the prior knowledge of the PI controller to reduce the search space and guide the RL exploration, which helps to reduce the exploration time and improve the sample efficiency. After a sufficient amount of data, the traditional agent finally catches the performance of the combined agent.

The combination of PI and RL, therefore, benefits from the advantages of both worlds. The PI controller provides a good indication of where the optimal control is located, while RL can refine it toward better controls.

\section{Conclusion}

In this work, we show that the combination of reinforcement learning and optimal control can be a powerful approach to designing near-optimal controllers for throttle valve systems. By adapting the recent area of Reinforcement Learning with Guides to this use case, we built a near-optimal controller that reduces the data requirement of traditional RL agents and overcomes the limitations of the PI controller. Our approach is validated on \startEW   three \stopEW valves with slightly different dynamics. 

This provides evidence that the integration of Control Theory techniques with RL can lead to significant improvements in data efficiency, making it a promising avenue for future research in other complex systems.

\section*{ACKNOWLEDGMENT}
The authors would like to thank Sophie CAPOBIANCO, Cheikh Saliou Fall NDIAYE and Adama DIABATE from Grenoble INP ENSE3 for their diligent efforts in realizing some of the experiments presented in this paper.

\addtolength{\textheight}{-0cm}   








\bibliography{Bib_TV.bib}
\bibliographystyle{IEEEtran}

\end{document}